\documentclass[twocolumn, showpacs, amsmath, amssymb, prd]{revtex4}
\usepackage{graphicx}
\usepackage{dcolumn}
\usepackage{bm}

\begin{document}

\title{Gauge-invariant momentum and angular momentum operators 
in quantum electrodynamics and chromodynamics}

\author{C. W. Wong$^1$}
\email[Corresponding author: ]{cwong@physics.ucla.edu}
\author{Fan Wang$^{2,3}$}
\author{W. M. Sun$^{2,3}$}
\author{X. F. L\"u$^3$} 

\affiliation{$^1$ Department of Physics and Astronomy, University of California, 
Los Angeles, CA 90095-1547, USA}
\affiliation{$^2$ Department of Physics and Joint Center of Particle, 
Nuclear Physics and Cosmology, Nanjing University,\\
and PMO, CAS, Nanjing 210093, People's Republic of China,}
\affiliation{$^3$ Kavli Inst. for Theor. Phys. China, CAS, Beijing 100190, 
People's Republic of China}

\begin{abstract}
Differences between vector potentials in different gauges contain no dynamics in 
both classical and quantum electrodynamics and chromodynamics. Consequently, 
once gauge invariance is established, results 
calculated in non-covariant gauges can be expected to agree with results obtained
in covariant gauges in all Lorentz frames. We show in particular that canonical
quantization in the Coulomb gauge can be used without giving up explicit gauge
invariance. Quantization in the Coulomb gauge is particularly simple because it 
involves only the two transverse photons/gluons present in all gauges. These
transverse photons/gluons reside on a 2-dimensional physical plane in momentum 
space perpendicular to the photon/gluon momentum {\bf k}. Explicit expressions 
are given for the basic momentum, spin and orbital angular momentum field 
operators of photons/gluons in the Coulomb gauge. Their properties are discussed 
in some detail. In particular, these field operators are shown to be more
complicated than the corresponding operators in quantum mechanics which they also
contain.
\end{abstract}

\maketitle

\section{Introduction}

In classical electrodynamics (CED), the electromagnetic (e.m.) fields are uniquely
determined by the Maxwell equations. The Maxwell theory contains a hidden 
Lorentz symmetry first uncovered by Einstein \cite{Einstein05} which can be 
displayed explicitly by using a vector potential. Unfortunately, the vector 
potential is not uniquely determined by the Maxwell equations. There are 
infinitely many vector potentials or gauges that generate the same e.m. fields
\cite{Jackson01}. This redundant gauge degree of 
freedom has to be removed by a choice of gauge before any actual calculation 
using the vector potential. It is obvious, however, that the e.m. fields generated 
by any two different vector potentials $A_n^\mu(x), n = 1,2,$ must be the same,
so that the difference $A_{\rm nd}^\mu(x) = A_2^\mu(x) - A_1^\mu(x)$ is a 
function of the spacetime location $x$ that carries no dynamical information 
whatsoever. Such a nondynamical (hence the subscript ``nd'') vector potential 
has the functional form $\partial^\mu \Lambda(x)$, where $\Lambda(x)$ is any 
twice differentiable local function of $x$, because the e.m. field 
$F_{\rm nd}^{\mu\nu} = \partial^\mu\partial^\nu \Lambda(x) 
- \partial^\nu \partial^\mu\Lambda(x)$ generated by it is identically zero.
In this paper, the word ``nondynamical'' will be used to describe a quantity 
that contains no dynamics.

Quantum ED (or QED) differs from CED by including the quantum properties of matter
and radiation. The charged particle is now an electron described by a relativistic
Dirac wave function $\psi$ spread out in space and accessible through a kind of 
square-root decomposition of the charge density: $\rho(x) = e\psi^\dag(x) \psi(x)$, 
where $e = -|e|$ is the electron's charge. The electron has the quantum momentum 
density distribution in space $\psi^\dag \hat{p}^\mu \psi$, where the quantum 
(or quantum mechanical) momentum operator $\hat{p}^\mu = \partial^\mu/i$ has to be 
placed between the two wave functions. This momentum density does not change if a 
global phase factor 
$\exp(i\lambda)$ with an $x$-independent phase $\lambda$ is added to $\psi$. 
On the other hand, if the phase is a twice differentiable local function of $x$, 
the momentum density changes to $\psi^\dag (\hat{p}^\mu + \partial^\mu\lambda) \psi$. 
Thus the expression is not invariant under a local phase transformation described 
by a phase function $\lambda(x)$ that contains no dynamics. 

In QED, the vector potential plays a role more fundamental than the e.m. fields 
because it has the Lorentz vector structure of spacetime itself. The Lorentz force 
law of CED shows that the charge-field interaction changes the point particle's 
canonical momentum $\hat{p}^\mu$ into the dynamical (usually called kinematical
or $mv$ \cite{Feynman65}) momentum $\hat{p}^\mu - eA^\mu(x)$. 
In QED too, dynamics now expressed as an
electron-photon ($e\gamma$) interaction can be included explicitly by using the 
dynamical momentum density distribution $\psi^\dag (\hat{p}^\mu - eA^\mu) \psi$ 
in space. Under a simultaneous gauge transformation of the vector potential and a 
local phase transformation of the wave function, the operator part of the expression 
is changed to $\hat{p}^\mu + \partial^\mu\lambda - e(A^\mu + \partial^\mu \Lambda)$.
The original expression is restored if the two changes are chosen to cancel each 
other perfectly: $\partial^\mu\lambda = e\partial^\mu\Lambda$ 
\cite{Fock27,Jackson01,Sakurai67,Cheng88}. This is the 
archetype of manifestly gauge invariant expressions usually used in QED, including 
the momentum and angular momentum (AM) densities proposed by Ji \cite{Ji97} and by 
Chen and Wang \cite{Chen97} in the nucleon spin problem. With manifest gauge 
invariance, one is assured that calculations made in any gauge give the same 
physical results as those obtained in any other gauge.

However, neither the nondynamical phase function $\lambda(x)$ nor the nondynamical 
gauge function $\Lambda(x)$ contains any $e\gamma$ interaction dynamics. The 
dynamics is already present in the Lorentz force law. We are only concerned 
with its invariance under a nondynamical gauge transformation, so that this 
gauge transformation itself does not add extra dynamics into the problem. Such 
a requirement has the same significance whether or not the $A$ that appears in 
the expression contains any dynamics. Moreover, the nondynamical expressions 
are closer to the expressions obtained directly from the QED Lagrangian. For 
these reasons, we have previously proposed leaving out the $e\gamma$ 
interaction altogether in constructing a gauge invariant version \cite{Chen08} of
the operators proposed by Jaffe and Manohar \cite{Jaffe90}, where the gluon
orbital AM and spin operators are separated. In contrast, Wakamatsu \cite{Waka10} 
has more recently advocated the dynamical version of these operators.

A somewhat more general conceptual scheme for realizing nondynamical gauge 
invariance works as follows: Let a general vector potential be written as 
$A^\mu = A^\mu_{\rm quan} + A^\mu_{\rm nd}$, where $A^\mu_{\rm quan}$ is the 
vector potential in the gauge where second quantization will be realized. 
Then the nondynamical momentum density 
$\psi^\dag (\hat{p}^\mu - eA^\mu_{\rm nd}) \psi$ is manifestly gauge invariant. 
Moreover, the nondynamical quantum momentum operator 
$\hat{p}^\mu - eA^\mu_{\rm nd}$ satisfies all the commutation relations of 
$\hat{p}^\mu$ alone. In contrast, different components of the dynamical momentum 
$\hat{p}^\mu - eA^\mu$ do not commute with one another \cite{Chen97}. Hence our 
nondynamical, gauge-invariant QED momentum operator is also a canonical momentum.

We note in passing that in this paper, we call the quantum operator 
$\hat{\bf p} = \bm{\nabla}/i$ a ``nondynamical'' operator in the sense that its 
eigenfunctions $\exp{(i{\bf k}\cdot{\bf x})}$ are momentum eigenfunctions describing 
a free particle of momentum {\bf k} in free space. It is technically a canonical
momentum. Feynman calls $\hat{\bf p}$ simply and unambiguously the ``p-momentum'' 
\cite{Feynman65}. However, it is sometimes also called perhaps confusingly a 
``dynamical momentum'' \cite{Feynman65}, a usage we do not follow. In this paper, 
a ``dynamical'' operator refers to one containing dynamics or interactions, unlike 
a ``nondynamical'' operator. 

The nondynamical phase factors considered here are functions only of spacetime 
locations. Their phase functions are obtained from a phase integral that is 
path-independent or integrable. There is another type of phase factors involving 
path-dependent or nonintegrable phase integrals that can add dynamics to the 
fermion field \cite{Dirac31, WuYang75,Sakurai67}. It is amusing to note that in the 
local gauge theory of interactions \cite{Cheng88}, local gauge invariance stated 
in terms of the nondynamical, path-independent phase factor is used to deduce 
the mathematical structure of the dynamical, path-dependent phase factor. 
However, the local gauge theory of interactions is not an immediate 
concern of this paper.

The purpose of this paper is to present and discuss the properties of the 
nondynamical, gauge-invariant operators in QED and QCD \cite{Chen08}. Our method 
is very flexible because it does not exclude the use of dynamical operators. 
On the other hand, the separation of the nondynamical momentum from the 
$e\gamma$ interaction seems to provide a more practical calculation scheme 
from the perspectives of experimental measurement, AM algebra and perturbation 
theory. Our presentation will be detailed and self-contained, in order to 
appeal to a wide readership. A historical note:  According to Jackson and Okun
\cite{Jackson01,Jackson99}, the Lorenz gauge, usually called the Lorentz
gauge and attributed to H. A. Lorentz, was first proposed by L. V. Lorenz in 1867. 

We shall begin in Sect. II with QED. Its natural habitat is the four-momentum space
where the Maxwell equations appear as coupled algebraic equations that can be solved 
analytically. Explicit formula for the vector potential will be given for a family 
of gauge choices that include the Lorenz and Coulomb gauges. These formulas will 
be needed to construct the more complicated operators in QCD. 

None of the vector potentials $A^\mu_n$ of the family is a Lorentz vector except 
those in covariant gauges such as the Lorenz gauge. (Note that $A^\mu$ is a 
Lorentz vector if it satisfies the covariant gauge condition $C_\mu A^\mu =$ 
Lorentz scalar, where $C_\mu$ is a known Lorentz vector such as $\partial_\mu$ or 
$x_\mu$.) Hence manifest Lorentz covariance and gauge invariance can coexist only 
for canonical quantization in a covariant gauge such as the Lorenz gauge. 
The trouble is that the covariant $A^\mu$ 
then generates four covariant photons: Among these photons, the longitudinal 
photon is unphysical and the time-like photon is virtual, never free of its 
source charge. Only the two remaining photons are transverse photons that 
can propagate in free space. 

In all non-covariant gauges, manifest Lorentz covariance of the vector 
potential is lost. Photon quantization must then proceed sequentially by first 
using a covariant gauge to ensure Lorentz covariance 
of the vector potential in different Lorentz frames. Only then can 
one work in a single Lorentz frame to construct manifestly gauge invariant 
operators. From this perspective, gauge invariance implies Lorentz covariance, 
but Lorentz covariance does not imply gauge invariance. That is, gauge invariance 
in one frame includes a covariant gauge where Lorentz covariance can be restored. 
On the other hand, Lorentz covariance in different frames alone does not involve 
gauge invariance at all.

The 3-vector potential is particularly simple in the Coulomb gauge where it 
contains only the 2-vector part 
$\bm{{\cal A}}_C  = \bm{{\cal A}}_\perp = \bm{{\cal A}}_{\rm phys}$ in momentum 
space perpendicular to the propagation vector {\bf k} of the light wave. This
2-vector potential is common to all the gauges in our chosen family of gauges.
It contains all the physics of the two transverse free photons. Consequently, 
we shall call this transverse plane the ``physical plane'' (hence the subscript 
``phys''.) Although $\bm{{\cal A}}_{\rm phys}$ is defined only after a Lorentz 
frame has been chosen, it nevertheless develops a hidden Lorentz symmetry under
the special circumstances of QED, namely that $\bm{{\cal A}}_{\rm phys}^2$ is a 
Lorentz scalar for all the $\bm{{\cal A}}_{\rm phys}$ of different Lorentz frames. 

The construction of momentum and AM operators in QED using 
Noether's theorem directly in terms of the vector potential as a field 
variable gives results where the nondynamical electron and photon terms are 
cleanly separated. Dynamical operators can next be introduced if desired by
adding a canceling action-reaction pair of $e\gamma$ interaction terms and 
associating one partner term with the electron operator and its canceling 
partner with the photon operator. 

Our nondynamical photon orbital AM and spin operators are well-defined 
gauge-invariant operators. Each operator has a well-defined component along the
photon momentum {\bf k} that describes the effect of 2-dimensional rotations 
about {\bf k}. They are therefore separately useful for problems with 
cylindrical symmetry about {\bf k}. The physical transverse photon states can be
expressed in terms of eigenstates of the total AM ${\bf J} = {\bf L} + {\bf S}$. 
We shall explain why the absence of longitudinal photon states in free space makes 
it impossible to observe the photon {\bf L} and {\bf S} separately and completely
in free space.

The extension of the above treatment to QCD is conceptually the same, but 
technical more complicated because of the nonlinear interactions among the 
vector potentials. We show why the family of gauge conditions used for CED can 
also be defined in CCD. Then an explicit perturbative solution in momentum space 
for the nondynamical part ${\cal A}^\mu_{\rm nd}$ can be obtained. 
(Note that ${\cal A}^\mu_{\rm pure}$ \cite{Chen08, Chen09, Fan09} is 
just ${\cal A}^\mu_{nd}$ for the special case where quantization is realized in 
the Coulomb gauge.) In particular, the vector potential ${\bf A}_\perp$ 
transverse to the gluon propagation direction {\bf k} is found to be 
gauge-independent, just like CED, at least for our perturbative solutions. 
We therefore do not use the conceptual method 
for QCD proposed in \cite{Chen08, Chen09, Fan09}, but construct 
QCD momentum and AM operators directly in complete analogy to the QED operators. 
Canonical quantization on the physical plane in the Coulomb gauge shows that the 
QCD operators, though nominally nondynamical like their QED counterparts from the
perspective of gauge transformations, nevertheless contain nonlinear interaction 
terms proportional to the color coupling constant $g$. Hence the complete removal 
of dynamical effects from the QCD operators seems impossible.

Sect. IV contains concluding remarks, where we touch upon the recent
criticism of our operators published by Ji \cite{Ji10}.

\section{Gauge invariant operators in QED}

We begin in Sect. A with classical ED. The properties of vector potentials 
and their explicit functional form in a certain family of gauge choices will be 
derived. In Sect. B, we construct gauge invariant nondynamical and dynamical 
momentum operators and briefly discuss their merits. Angular momentum operators
are constructed in Sect. C. Photon AM operators are examined in detail in
Sect. D, and their eigenstates of good total AM are obtained in Sect. E. 
The units $\hbar = c = 1$ are used.

\subsection{Vector potentials in classical ED}
\label{AinCED}

In a chosen Lorentz frame, the vector potential has a particularly simple 
structure in the 4-dimensional Fourier space $k^\mu = (\omega, {\bf k})$ 
\cite{CT89, Wong09}: 
${\cal A}^\mu(k) = ({\cal A}^0, \bm{{\cal A}}_\parallel + \bm{{\cal A}}_\perp)$. 
It contains a two-dimensional part $\bm{{\cal A}}_\perp$ transverse to the 
propagation vector {\bf k} and known to be gauge independent because it 
is completely determined by the gauge-independent magnetic induction:
\begin{eqnarray}
\bm{{\cal A}}_\perp  
= \frac{i}{|{\bf k}|^2} {\bf k} \bm{\times {\cal B}_\perp}. 
\label{Aperp}
\end{eqnarray}
Once known, $\bm{{\cal A}}_\perp$ determines $\bm{{\cal E}}_\perp$ from the 
Faraday effect:
\begin{eqnarray}
\bm{{\cal E}}_\perp  =  i\omega \bm{{\cal A}}_\perp. 
\label{Eperp}
\end{eqnarray}
There is now only one e.m. field 
$\bm{{\cal E}}_\parallel = {\cal E}_\parallel {\bf e}_{\bf k}$ left to be related 
to the two leftover components ${\cal A}^0$ and ${\cal A}_\parallel$. 
The remaining relation between them is usually written as
\begin{eqnarray}
{\cal E}_\parallel 
= -i|{\bf k}|{\cal A}^0 + i\omega{\cal A}_\parallel,
\label{EparallelPhiA}
\end{eqnarray}
showing that there are infinitely 
many choices of ${\cal A}^0$ and ${\cal A}_\parallel$ for the same 
${\cal E}_\parallel$. This is the redundant gauge degree of freedom. 

We are interested in a family of vector potentials defined {\it in any
Lorentz frame} by the following real linear relation, to be called the 
generalized velocity gauge condition: 
\begin{eqnarray}
{\cal A}^{(vg)}_\parallel = s\alpha\frac{\omega}{|{\bf k}|} {\cal A}^{0(vg)},
\quad s = \pm 1,
\label{gaugeCond}
\end{eqnarray}
where the gauge parameter $\alpha = 1/v_g^2$ is expressed in terms of a gauge 
velocity $v_g$ that is a Lorentz scalar. The choice $s = 1$ has been studied in 
\cite{Yang05, Jackson02, Wong09}. We are here extending the family by including 
the subfamily with the signature $s = -1$. (If this $s = -1$ signature factor is 
absorbed into $v_g^2$, the gauge velocity will become purely imaginary.) 
The Coulomb gauge $\alpha_C = 0$ appears in both subfamilies. The gauge independence 
of ${\cal E}_\parallel$ in Eq. (\ref{EparallelPhiA}) requires that \cite{Wong09}
\begin{eqnarray}
{\cal A}^{0(vg)} = \frac{|{\bf k}|^2}{|{\bf k}|^2 - s\alpha\omega^2} 
{\cal A}^0_C.
\label{A0vg}
\end{eqnarray}
That is, Eq. (\ref{A0vg}) is a solution of the simultaneous linear equations
(\ref{EparallelPhiA}, \ref{gaugeCond}) for the unknowns ${\cal A}^{0(vg)}$ and
${\cal A}^{(vg)}_\parallel$ in any Lorentz frame. 

Among these gauges, the Coulomb gauge provides the 
most economical description because $\bm{{\cal A}}_{C\parallel} = 0$ and
${\cal A}_C^0 = {\cal E}_\parallel/(i\omega)$. 
That is, $\bm{{\cal A}}_C = \bm{{\cal A}}_\perp = \bm{{\cal A}}_{\rm phys}$  
contains only the gauge-independent and physics-containing part common 
to all gauges that resides on the {\it physical} plane. We shall therefore call 
the Coulomb gauge the physical gauge. The vector potential in any gauge is 
then $A^\mu = A^\mu_{\rm phys} + A^\mu_{\rm pure}$, where the subscript 
``pure'' refers to the nondynamical pure-gauge part. 

It is interesting to note here that in QED, $\bm{{\cal A}}_\perp$ on the physical
plane develops a hidden Lorentz symmetry: ${\cal A}_\perp^2$ is a Lorentz scalar 
for one-photon states. This symmetry appears because in QED, $A^\mu$ in a covariant 
gauge is to be quantized into a one-photon operator. Consistent with this objective, 
its Fourier space $(\omega, {\bf k})$ is specialized into the 4-momentum space of 
a massless photon satisfying the condition $\omega = |{\bf k}|$. For example, 
the Lorenz gauge condition (\ref{gaugeCond}) then reads 
\begin{eqnarray}
{\cal A}_{L\parallel} = {\cal A}_L^0.
\label{LorGaugeCond}
\end{eqnarray}
Then the Lorentz scalar ${\cal A}_L^2$ of the vector 
potential in the Lorenz gauge is just
\begin{eqnarray}
{\cal A}_L^2 &=& {\cal A}_L \cdot {\cal A}_L = {\cal A}_L^\mu {\cal A}_{L\mu} 
\nonumber \\
&=& \left( {\cal A}_L^0 \right)^2
- \left[ \left({\cal A}_L^0 \right)^2 + {\cal A}_\perp^2 \right] 
= - {\cal A}_\perp^2.
\label{LorScalar}
\end{eqnarray}
[We use the metric $g^{\mu\nu} = {\rm diag}(1, -1, -1, -1)$ for covariant 
quantities. For 3-dimensional vectors, however, we do not use a tensor notation 
so that spatial indices have the same meaning as superscripts or subscripts. 
Their location is determined only by the readability of the expression.] 
Under Lorentz transformations, both the orientation of the physical plane as 
defined by the propagation direction ${\bf e}_{\bf k}$ and the actual direction 
of $\bm{{\cal A}}_\perp$ on the physical plane can change, but not the length 
${\cal A}_\perp$ contained in ${\cal A}_L^\mu$ that is common to all the 
gauges under consideration. 

Furthermore, starting with the Lorentz vector potential $A^\mu_L$ in a new 
Lorentz frame, we can gauge transform in a standard way to any other velocity 
gauge \cite{Wong09}. The resulting vector potential in this new gauge in the 
new Lorentz frame must satisfy the gauge condition of Eq. (\ref{gaugeCond}), by 
definition. In the Coulomb-gauge, for example, ${\cal A}_{C\parallel} = 0$ and 
therefore the Coulomb-gauge condition ${\bf k} \cdot \bm{{\cal A}} = 0$ are 
guaranteed in the new Lorentz frame. From this property in classical ED, one can 
expect that QED formulated in any velocity gauge gives the same physical results
even though the vector potential is in general not a Lorentz vector. For an 
explicit demonstration of the Lorentz invariance of QED in the Coulomb gauge, 
see \cite{Zumino60, Lee81}. 

In the $s = -1$ subfamily of velocity gauges, the choice $\alpha = 1$ gives
rise to the covariant light-cone (lc) gauge where 
${\cal A}_{lc\parallel} = - {\cal A}_{lc}^0$ for one-photon states. Consequently, 
${\cal A}_{lc}^2 = - {\cal A}_\perp^2$ holds also.

Returning to CED, we note that another hidden Lorentz symmetry can be said to 
exist in the ${\cal A}^\mu_{\rm nd}$ between any two gauges, namely it generates 
no e.m. field ($F^{\mu\nu}_{\rm nd} = 0$) in any Lorentz frame even though it is 
not a Lorentz vector. 

The family of vector potentials even in one gauge on one Lorentz frame is very 
large, for the gauge condition (\ref{gaugeCond}) can be written more generally 
as 
\begin{eqnarray}
{\cal A}^{(vg)}_\parallel + {\cal \chi} = 
s\alpha\frac{\omega}{|{\bf k}|} \left( {\cal A}^{0(vg)} 
+ \frac{\omega}{|{\bf k}|} {\cal \chi} \right),
\label{enGaugeCond}
\end{eqnarray}
where ${\cal \chi}$, a solution of the ``wave'' equation
\begin{eqnarray}
(|{\bf k}|^2 - s\alpha \omega^2) {\cal \chi} = 0, 
\label{chiWaveEquation}
\end{eqnarray}
defines a subfamily of {\it restricted} gauge transformations in the same gauge.
The gauge velocity $v_g$ contained in the gauge parameter $\alpha = 1/v_g^2$ 
is the {\it nonphysical} wave velocity appearing in Eq. (\ref{chiWaveEquation}). 

Because the two components ${\cal A}^{0(vg)}$ and  ${\cal A}^{(vg)}_\parallel$ 
are described by a nonphysical wave velocity $v_g$, the resulting 
${\cal A}^{\mu(vg)}$ is not a Lorentz vector except when $v_g = 1$, in the Lorenz 
gauge when $s = 1$ and in the covariant light-cone gauge when $s = -1$. However, 
The two remaining components $\bm{{\cal A}}_\perp$ on the physical plane are 
gauge independent and satisfy physical wave equations involving $c = 1$. This is 
the mathematical reason why $\bm{{\cal A}}_\perp$ contains hidden Lorentz symmetry 
even though it is not a Lorentz 4-vector.

The Lorenz and covariant light-cone gauges are also members of the generalized 
covariant gauges satisfying the 
gauge condition $C_\mu A^\mu =$ Lorentz scalar, where $A^\mu$ is a Lorentz vector: 
Here $C_\mu$ is a known Lorentz vector and the word ``generalized'' refers to the 
use of a Lorentz scalar that is not zero. Most covariant gauges of practical 
importance use a zero Lorentz scalar. These include the Lorenz gauge 
[$C_\mu = k_\mu = (\omega, - {\bf k})$], the covariant light-cone 
(lc) gauge [$C_\mu = \tilde{k}_\mu = (\omega, {\bf k})$], and the 
relativistic Poincar\'e gauge [$C_\mu = x_\mu = (t, -{\bf x})$]. Covariant gauges 
like the relativistic Poincar\'e gauge are defined in spacetime where the physical 
plane in the Fourier space {\bf k} cannot be isolated explicitly. That is, 
only the Fourier transform $\bm{{\cal A}}_\perp$ lies on a 2-dimensional plane in 
the Fourier space {\bf k}.

The relation between ${\cal A}_\parallel$ and ${\cal A}^0$ can also be nonlinear:
${\cal A}_\parallel = f({\cal A}^0)$, where $f(x)$ is a nonlinear function of $x$.
The main complication of such a choice is that the gauge independence condition
of Eq. (\ref{EparallelPhiA}) in general has multiple roots. 

Last but not least, the classical e.m. momentum and AM densities
can be written in the following useful differential triple vector product forms
when expressed in terms of the vector potential {\bf A}:
\begin{eqnarray}
{\bf E} \times (\bm{\nabla}\times{\bf A}) &=& E^j \bm{\nabla} A^j 
+ (\bm{\nabla}\cdot{\bf E}){\bf A} \nonumber \\
&& - \;\nabla^j(E^j{\bf A}), 
\label{momDenID} \\
{\bf x} \times [{\bf E} \times (\bm{\nabla}\times{\bf A})] 
&=& E^j ({\bf x}\times\bm{\nabla}) A^j \nonumber \\
&& + \; (\bm{\nabla}\cdot{\bf E})({\bf x}\times{\bf A}) 
 +  {\bf E} \times {\bf A}  \nonumber \\
&&  - \; \nabla^j[E^j({\bf x}\times{\bf A})].
\label{angMomDenID}
\end{eqnarray}
Equation (\ref{angMomDenID}) is 
easily derived from Eq. (\ref{momDenID}). These elementary classical 
expressions already contain the differential operator and gauge-dependent 
substructures we shall find in QED. Most presentations of the subject use
these identities to change the classical expressions directly to their QED
relations. We shall work directly with the QED Lagrangian, however, in order to 
provide a different perspective.

\subsection{Gauge invariant nondynamical and dynamical momentum operators in QED}
\label{QEDmom}

The QED Lagrangian density is
\begin{eqnarray}
\mathcal{L} &=&
\bar{\psi}[i\gamma^{\mu}(\partial_{\mu} - ieA_{\mu})-m]\psi
-\frac{1}{4}F_{\mu\nu}F^{\mu\nu}, \nonumber\\
F_{\mu\nu} &=& \partial_{\mu}A_{\nu}-\partial_{\nu}A_{\mu},
\label{QEDLagDen}
\end{eqnarray}
where the field variables are $\psi, \bar{\psi}$ and $A_{\mu}$. According to 
Noether's theorem, the invariance of ${\cal L}$ under translation generates a 
conserved energy-momentum tensor $T^{\mu\nu}$. The $e$ and $\gamma$ momentum 
densities
\begin{eqnarray}
T^{0n} &=& T_e^{0n} + T_\gamma^{0n} \nonumber \\
&=& \psi^{\dag}(\nabla^n/i)\psi + 
E^j\nabla^nA^j,
\label{QEDmomDen}
\end{eqnarray}
contain no direct contribution from the $e\gamma$ interaction term in ${\cal L}$. 
Here $n, j$ are Cartesian coordinate labels in (3-dimensional) space. 
Repeated indices imply summation. 

Since the classical Maxwell theory is Lorentz covariant, we may next choose a 
Lorentz frame and work with the momentum operator in 3-dimensional space, 
provided that gauge invariance is verified afterwards for all physical operators. 
This is the price paid for bypassing the gauge independent {\bf B} field. 
The total QED momentum operator is then
\begin{eqnarray}
{\bf P} &=& {\bf P}^e + {\bf P}^\gamma  \nonumber \\
&=& \int \psi^{\dag} \hat{\bf p} \psi \,d^3x
+ \int [E^j (\bm{\nabla} A^j)]_{\rm Herm} \,d^3x, 
\label{Ptot}
\end{eqnarray}
where $\hat{\bf p} = \bm{\nabla}/i$ is the quantum canonical momentum, and 
\begin{eqnarray}
[E^j (\bm{\nabla} A^j)]_{\rm Herm} 
= \frac{1}{2}[E^j ( \bm{\nabla}A^j ) +  ( \bm{\nabla}A^j )E^j ]
\label{Herm}
\end{eqnarray}
is Hermitian. For notational simplicity, the subscript ``Herm'' will not be 
shown henceforth. The photon momentum has a gauge-invariant physical part and 
a nondynamical, gauge-dependent or pure-gauge part:
\begin{eqnarray}
{\bf P}^\gamma &=& {\bf P}^\gamma_{\rm phys} + {\bf P}^\gamma_{\rm pure}  
\nonumber \\
&=& \int E^j_\perp ( \bm{\nabla} A^j_\perp ) \,d^3x, 
+ \int E_\parallel ( \bm{\nabla} A_\parallel )\,d^3x. 
\label{Pgam}
\end{eqnarray}
Note that four terms appear on writing 
${\bf E} = {\bf E}_\perp + {\bf E}_\parallel$ and 
${\bf A} = {\bf A}_\perp + {\bf A}_\parallel$. The two cross terms not shown 
vanish separately: Their structures are more transparent when their space 
integrals are re-written as momentum integrals. Each integral then involves 
perpendicular vector fields $\bm{{\cal E}}^*_\perp$ and $\bm{{\cal A}}_\parallel$ 
(or $\bm{{\cal E}}^*_\parallel$ and $\bm{{\cal A}}_\perp$) that form a zero 
scalar product when $\bm{\nabla} \rightarrow i{\bf k}$ is moved out of their way. 

The gauge-dependent term ${\bf P}^\gamma_{\rm pure}$ can be moved to the electron
momentum operator to make the latter manifestly gauge invariant. This is done by
first express ${\bf P}^\gamma_{\rm pure}$ in a familiar form with the help of the 
differential vector identity (\ref{momDenID}) used with 
${\bf E} = {\bf E}_\parallel$ and $\bm{\nabla}\times{\bf A}_\parallel = 0$:
\begin{eqnarray}
0 = E_\parallel ( \bm{\nabla} A_\parallel) 
+ (\bm{\nabla} \cdot {\bf E}_\parallel) {\bf A}_\parallel 
- \nabla^j (E^j_\parallel{\bf A}_\parallel).
\label{vectorID}
\end{eqnarray}
The volume integral over space of the last term on the right can be changed into 
a surface integral ``at infinity'' by using the Gauss theorem 
($\int d^3x \nabla^j \rightarrow \int_S d\sigma^j$). The surface integral vanishes 
if the fields are vanishingly small on the large surface at infinity. Then
\begin{eqnarray}
{\bf P}^\gamma_{\rm pure} = - \int \psi^{\dag} e{\bf A}_\parallel \psi \,d^3x,
\label{PgamPure}
\end{eqnarray}
where use has been made of the Gauss law 
$\bm{\nabla} \cdot {\bf E}_\parallel = e\psi^{\dag}\psi$. Added to ${\bf P}^e$,
this nondynamical longitudinal momentum gives manifest gauge invariance to the
nondynamical electron momentum
\begin{eqnarray}
{\bf P}^e_{\rm nd} &=& {\bf P}^e + {\bf P}^\gamma_{\rm pure}  \nonumber \\
&=& \int \psi^{\dag} (\hat{\bf p} - e{\bf A}_\parallel)\psi \,d^3x.
\label{PeGauInv}
\end{eqnarray}
Once gauge invariance is assured, we can work in the Coulomb gauge where the 
electron momentum is just ${\bf P}^e$. Note that the complete gauge invariant
4-momentum density is $\psi^{\dag}(p^\mu - eA^\mu_{\rm pure})\psi$, where 
in momentum space, both $k^\mu$ and ${\cal A}^\mu_{\rm pure}$ have only
nonzero $\mu = 0, 3$ components. Their identical mathematical structures explain 
why they appear naturally together.

To get Ji's dynamical electron momentum operator:
\begin{eqnarray}
{\bf P}^e_{\rm dyn} &=& \int \psi^{\dag} (\hat{\bf p} - e{\bf A})\psi \,d^3x
\nonumber \\
&=& {\bf P}^e_{\rm nd} - {\bf P}^{e\gamma}_\perp.
\label{PeDyn}
\end{eqnarray}
we have to add to our nondynamical ${\bf P}^e_{\rm nd}$ a transverse momentum 
$-{\bf P}^{e\gamma}_\perp$ from the $e\gamma$ interaction, where 
\begin{eqnarray}
{\bf P}^{e\gamma}_\perp = \int \psi^{\dag} e{\bf A}_\perp \psi \,d^3x.
\label{PeGamInt}
\end{eqnarray}
The canceling partner ${\bf P}^{e\gamma}_\perp$ of this action-reaction pair of 
terms is then added to the photon momentum ${\bf P}^\gamma_{\rm phys}$ to 
give a dynamical photon momentum operator
\begin{eqnarray} 
{\bf P}^\gamma_{\rm dyn} = {\bf P}^\gamma_{\rm phys} 
+ {\bf P}^{e\gamma}_\perp.
\label{PgamDyn}
\end{eqnarray}
Such a re-definition does not change the total momentum of the $e\gamma$ system. 

Is such a change desirable? First note that after the ${\bf P}^{e\gamma}_\perp$ 
term is added, the photon momentum ${\bf P}^\gamma_{\rm dyn}$ is no longer parallel 
to its free-space momentum {\bf k}. That is, the interacting photon is off shell. 
This unusual feature must be kept in mind if the modified operator is used. 

Secondly, the two types (nondynamical and dynamical) of electron momentum 
operators satisfy different commutation relations. Both operators satisfy the 
momentum-position commutation relations
\begin{eqnarray}
\left[ \hat{p}^\mu - eA^\mu(x), x_\nu \right] = -i\delta^\mu_\nu,
\label{canCommRel}
\end{eqnarray}
if $A^\mu(x)$ is momentum independent. However, between the momentum 
components themselves
\begin{eqnarray}
\left[ \hat{p}^\mu - eA^\mu(x), \hat{p}^\nu - eA^\nu(x) \right] = ieF^{\mu\nu},
\label{MomCommRels}
\end{eqnarray}
the commutators vanish only for the field-free $A^\nu_{\rm nd}(x)$ that appears in 
the  nondynamical momentum ${\bf P}^e_{\rm nd}$. In contrast, not all three 
components of the dynamical momentum ${\bf P}^e_{\rm dyn}$ are simultaneously 
observable \cite{Chen97}. These results remain unchanged in QCD.

Thirdly, an experimental momentum measurement of the $e$ or $\gamma$ bound in
an atomic state requires the use of an external probe to eject the measured 
particle from the atom and put it on the energy shell for detection at a distant 
detector. The detector cannot tell the past dynamical history of {\it one} 
detected free particle. It is only after extended measurements of suitable 
distribution functions when the past dynamical contents of the detected free 
particles can be deduced. In other words, unlike classical particle dynamics, 
the dynamics contained in a quantum wave function is not localized at 
{\it one point} in space but is spread out all over space in a probability 
density distribution. 

Theoretical analysis requires calculations of relevant transition matrix 
elements where the full dynamics contained in the atomic states can in principle 
be included through their wave functions. In practice, at some stage one is 
often forced to use perturbation theory where the unperturbed states are often 
constructed from free-space operators. After that, 
the $e\gamma$ interaction is included explicitly order by order. In covariant 
perturbation theory, for example, internal as well as external lines are labeled 
by free-space 4-momenta. All $e\gamma$ interactions appear as vertices where 
free-space energy-momentum conservation is imposed. The propagators describing 
internal lines also involve these free-space momenta without any interaction. 
Hence the nondynamical canonical momentum operators are well matched to these 
theoretical and experimental circumstances.

The interaction momentum term ${\bf P}^{e\gamma}_\perp$ in the dynamical $e$ and 
$\gamma$ momentum operators involve both particles. If we measure only one of the 
two particles, we do not know if the last vertex the particle comes from is not 
from other dynamical correlations in the atomic states rather than from this 
interaction momentum operator. If we measure both particles, we do not know if 
they actually come from the same interaction vertex, a knowledge that would 
qualify that vertex as the interaction momentum operator of interest. Moreover, 
perturbation theory generally does not calculate both members of a canceling 
action-reaction pair of terms, but simply set their total contribution to zero, 
as is done explicitly in the nondynamical canonical method.

Thus it appears that our nondynamical momenta are simpler and perhaps 
more natural and more practical than the dynamical operators advocated 
by Ji \cite{Ji97} and Wakamatsu \cite{Waka10}.

\subsection{Angular momentum operators in QED}

The construction of gauge invariant AM operators in QED proceeds in the same way as 
for gauge invariant momentum operators of the last subsection. Invariance of the QED 
Lagrangian density ${\cal L}$ of Eq. (\ref{QEDLagDen}) under rotation gives rise to 
a certain conserved moment tensor $M^{\alpha\beta\gamma}$ whose spatial component 
$M^{0bc}$ integrates to the three-dimensional total AM \cite{Jaffe90}
\begin{eqnarray}
{\bf J} &=& {\bf S}^e + {\bf L}^e + {\bf S}^\gamma + {\bf L}^\gamma \nonumber \\
&=& \int \psi^{\dag} (\bm{\Sigma}/2) \psi \,d^3x
+ \int \psi^{\dag} \hat{\bf L} \psi \,d^3x \nonumber \\
&& + \; \int {\bf E} \times {\bf A} \,d^3x 
+ \int  E^j (i\hat{\bf L} A^j) \,d^3x, 
\label{Jtot}
\end{eqnarray}
where $\Sigma^j = (i/2)\epsilon_{jmn}\gamma^m\gamma^n$ and 
$\hat{\bf L} = {\bf x} \times \hat{\bf p}$ is the quantum orbital AM 
momentum operator. The photon operators can be expressed as (with details given 
in the Appendix)  
\begin{eqnarray}
{\bf S}^\gamma + {\bf L}^\gamma  
&=& {\bf S}^\gamma_\parallel + {\bf L}^\gamma_{\rm phys} 
+ {\bf L}^\gamma_{\rm pure}, \nonumber \\
{\bf S}^\gamma_\parallel &=&  \int {\bf E}_\perp \times {\bf A}_\perp \,d^3x, 
\nonumber \\
{\bf L}^\gamma_{\rm phys} &=& \int  E^j_\perp (i\hat{\bf L} A^j_\perp) \,d^3x,
\nonumber \\
{\bf L}^\gamma_{\rm pure} &=& 
-\int \psi^{\dag} ({\bf x} \times e{\bf A}_\parallel) \psi \,d^3x.
\label{Jgam}
\end{eqnarray}
We get two physical, gauge-invariant terms ${\bf S}^\gamma_\parallel$
and ${\bf L}^\gamma_{\rm phys}$, and a nondynamical gauge-dependent term 
${\bf L}^\gamma_{\rm pure}$ to be added to the electron orbital AM to make it 
manifestly gauge invariant: 
\begin{eqnarray}
{\bf L}^e_{\rm nd} 
&=& {\bf L}^e + {\bf L}^\gamma_{\rm pure}  \nonumber \\
&=& \int \psi^{\dag} [{\bf x} \times (\hat{\bf p} - e{\bf A}_\parallel)] \psi \,d^3x. 
\label{eLnd}
\end{eqnarray}
The separation of ${\bf S}^\gamma_\parallel$ from ${\bf L}^\gamma_{\rm phys}$
has actually been known for many years (see, for example, \cite{Gottfried66, 
Jackson75, CT89}), as well as the implicit gauge independence of
Coulomb-gauge quantization \cite{Gottfried66}. 

Ji's dynamical electron AM operator
\begin{eqnarray}
{\bf L}^e_{\rm dyn} &=& \int \psi^{\dag} 
[{\bf x} \times (\hat{\bf p} - e{\bf A})\psi] \,d^3x
\nonumber \\
&=& {\bf L}^e_{\rm nd} - {\bf L}^{e\gamma}.
\label{eLdyn}
\end{eqnarray}
is obtained by adding to our nondynamical ${\bf L}^e_{\rm nd}$ a term 
$-{\bf L}^{e\gamma}$ that comes from the $e\gamma$ interaction, where 
\begin{eqnarray}
{\bf L}^{e\gamma} = \int \psi^{\dag} ({\bf x} \times e{\bf A}_\perp) \psi \,d^3x.
\label{LeGam}
\end{eqnarray}
The canceling partner ${\bf L}^{e\gamma}$ of this action-reaction pair of 
terms is then added to the photon momentum to give a dynamical photon AM operator
\begin{eqnarray} 
{\bf L}^\gamma_{\rm dyn} = {\bf L}^\gamma_{\rm phys} + {\bf L}^{e\gamma}.
\label{LgamDyn}
\end{eqnarray}

Equation (\ref{Jgam}) shows that in momentum space, the spin operator
$\bm{{\cal S}}^\gamma_\parallel  = {\cal S}^\gamma_\parallel {\bf e}_{\bf k}$  
is built up entirely of 2-dimensional photon fields on the physical plane. 
This {\it intrinsic} photon structure contains the helicity operator 
${\cal S}^\gamma_\parallel$ whose two eigenstates 
$|{\bf e}_\pm\rangle = \mp |({\bf e}_1 \pm i{\bf e}_2)/\sqrt{2}\rangle$ 
lie on the physical plane. The helicity 0 spin state $|{\bf e}_3\rangle$ parallel
to {\bf k} is absent because the two transverse spin operators on 
the plane that can generate it are absent, i.e., $\bm{{\cal S}}^\gamma_\perp = 0$. 
These properties are caused by the fact that the massless photon has no rest frame. 
Only the realizable nonzero photon helicities are physically observable.

On the other hand, the photon orbital AM ${\bf L}^\gamma$ has all three intact 
components for any choice of the quantization axis ${\bf e}_z$. It is an 
{\it external} 3-dimensional operator acting on the photon fields whose
directions are restricted to the physical plane. Normally a gauge invariant 
quantity like ${\bf L}^\gamma$ can be expected to be physically observable. 
We shall show in the next subsection, however, that with all longitudinal 
vector spherical harmonics perpendicular to the physical plane absent, the 
orbital AM of a free photon is only partially observable from the perspective of 
quantum measurement.

\subsection{Free photon operators and states}
\label{CanonQuantization}

The AM properties of a free photon become more transparent after canonical 
quantization in the Coulomb gauge \cite{Gottfried66, Sakurai67, Lee81, CT89, 
VEN94, Mandel95, Greiner96}, when the vector potential takes the simple form
\begin{eqnarray} 
{\bf A}_\perp (x) = \sum_{\bf k} \sqrt{\frac{1}{2\omega V}}
\left( {\bf a}_{\bf k} e^{-ik \cdot x} 
+ {\bf a}^\dag_{\bf k} e^{ik \cdot x} \right).  
\label{quantAperp}
\end{eqnarray}
Here $x^\mu = (t, {\bf x})$, and the 3-momentum state labels {\bf k} may take a  
non-Cartesian form to match the geometry of the quantization volume $V$. For the 
proper treatment of rotational properties, for example, it is best to use 
a spherical box. Then the AM eigenvalues contained in the {\bf k} labels remain 
discrete as $V \rightarrow \infty$. (The same generic {\bf k} notation can be used 
for a cubic or cylindrical box.) The transverse operators 
\begin{eqnarray} 
{\bf a}_{\bf k}  = \sum_{\lambda=\pm} a_{{\bf k}\lambda} {\bf e}_\lambda  
\label{aaDag}
\end{eqnarray}
and ${\bf a}^\dag_{\bf k}$ destroy and create a single photon, respectively, 
in the Heisenberg (time-independent) representation. The time dependence
appears solely in the four scalar product 
$k \cdot x = \omega t - {\bf k} \cdot {\bf x}$, with $\omega = |{\bf k}|$ for 
one-photon states. 

The fields ${\bf E}_\perp(x)$ and ${\bf B}(x)$ for free photons can be calculated 
directly from ${\bf A}_\perp(x)$. Other operators can then be reduced easily to second 
quantized form. Thus the Hamiltonian, without the divergent zero-point energy term, 
and the momentum can be expressed in terms of number operators:
\begin{eqnarray} 
H^\gamma &=& \frac{1}{2} \int \left( {\bf E}^2_\perp + {\bf B}^2 \right) d^3x
= \sum_{\bf k} \omega N_{\bf k}, \\
{\bf P}^\gamma_\parallel &=& \sum_{\bf k} {\bf k} N_{\bf k} 
= \sum_{\bf k} {\bf k} (N_{{\bf k}+} +  N_{{\bf k}-}), \\
N_{\bf k} &=& {\bf a}^\dag_{\bf k} \cdot {\bf a}_{\bf k} 
= N_{{\bf k}+} +  N_{{\bf k}-}.
\label{HgamPgam}
\end{eqnarray}

The spin operator shows a characteristic spinning or vector product structure 
that can be simplified into two distinct final forms, a helicity form if 
${\bf e}_z = {\bf e}_{\bf k}$ and a 3-dimensional form if ${\bf e}_z$ is 
any direction:
\begin{eqnarray} 
{\bf S}^\gamma_\parallel &=& 
-i\sum_{\bf k} {\bf a}^\dag_{\bf k} \times {\bf a}_{\bf k} 
= \sum_{\bf k} {\bf e}_{\bf k} (N_{{\bf k}+} -  N_{{\bf k}-}), \\
&=&  \sum_{\bf k} \sum_{m,n} ( {\bf a}^\dag_{\bf k} )_m
\hat{\bf S}_{mn} ( {\bf a}_{\bf k} )_n
= \sum_{\bf k} {\bf a}^\dag_{\bf k} 
\hat{\bf S}{\bf a}_{\bf k}.
\label{Sgam}
\end{eqnarray}
Here $m, n = 1, 2, 3$ are Cartesian indices and 
$( {\bf a}_{\bf k} )_n = {\bf a}_{\bf k} \cdot {\bf e}_n$ is a Cartesian component 
in the chosen frame. The 3-dimensional form contains an expected quantum spin 
vector for spin 1 particles, 
\begin{eqnarray} 
\hat{\bf S} = \sum_j {\bf e}_j\hat{S}_j , \quad {\rm where} \quad
(\hat{S}_j)_{mn} = -i \epsilon_{jmn},
\label{quantS}
\end{eqnarray}
that is a $3 \times 3$ matrix in spin space with Cartesian components $\hat{S}_j$
that are $3\times 3$ Hermitian matrices defined by the Levi-Civita symbols from 
the cross product. It satisfies the standard quantum AM algebra 
$\hat{\bf S} \times \hat{\bf S} = i\hat{\bf S}$. 
In this matrix notation, the 3-dimensional vectors 
${\bf a}_{\bf k},{\bf a}_{\bf k}^\dag$ in the last expression of Eq. 
(\ref{Sgam}) are column and row vectors, respectively, in spin space.

Finally, the orbital AM operator has the same general structure: 
\begin{eqnarray} 
{\bf L}^\gamma = \sum_{\bf k} \sum_{\lambda=\pm} 
a^\dag_{{\bf k}\lambda} \hat{\bf L}\, a_{{\bf k}\lambda}
=  \sum_{\bf k} {\bf a}^\dag_{\bf k} \hat{\bf L} {\bf a}_{\bf k}. 
\label{Lgam}
\end{eqnarray}
Note the subtle difference between the expression for ${\bf L}^\gamma$ where
$\hat{\bf L} = - {\bf k} \times \hat{\bf x}$ acts on the components  
$a_{{\bf k}\lambda}$ of the 2-vector ${\bf a}_{\bf k}$ on the physical plane, and 
${\bf S}^\gamma$ where $\hat{\bf S}$ acts on the unit vectors ${\bf e}_j$ 
contained in the same vector ${\bf a}_{\bf k}$. The last term in Eq. (\ref{Lgam}) 
can also be considered a matrix expression with $\hat{\bf L}$ a non-matrix quantity, 
or a diagonal matrix in spin space if an optional $3 \times 3$ identity matrix is 
added. The 3-dimensional forms for both ${\bf L}^\gamma$ and ${\bf S}^\gamma$ are 
needed in real space where the physical plane cannot in general be isolated explicitly.

Three of the operators, $H^\gamma, {\bf P}^\gamma$ and ${\bf S}^\gamma_\parallel$, 
are functions only of the number operators that commute among themselves. Hence 
the $n$-photon number state $|\{n_i\}\rangle$ is also the simultaneous eigenstate 
of these operators with the respective eigenvalues
\begin{eqnarray} 
E = \sum^n_{i=1} \omega_i, \quad
{\bf K} = \sum_i {\bf k}_i, \quad
\bm{\Lambda} &=& \sum_i \lambda_i {\bf e}_i,
\label{gamEigenvalues}
\end{eqnarray}
where ${\bf e}_i = {\bf e}_{{\bf k}_i}$ of the $i$-th free photon. In particular, 
the one-photon state of momentum {\bf k} can be written in the helicity form 
\cite{Gottfried66}
\begin{eqnarray} 
|\alpha = {\bf k}\lambda\rangle = a^{\dag}_\alpha |0\rangle.
\label{photonState}
\end{eqnarray}
This photon state carries only 4 labels or constants of motion: $k^j, \,j=1,2,3$ 
(including $\omega = |{\bf k}|$), and a single helicity. The Cartesian components 
of the spin vector $\bm{{\cal S}}^\gamma_\parallel$ in momentum space commute 
among themselves \cite{VEN94},
\begin{eqnarray} 
[ {\cal S}^\gamma_{\parallel m}, {\cal S}^\gamma_{\parallel n}] = 0,
\label{SparaCommRels}
\end{eqnarray}
where  ${\cal S}^\gamma_{\parallel m} = {\cal S}^\gamma_\parallel 
({\bf e}_{\bf k} \cdot {\bf e}_m)$. $\bm{{\cal S}}^\gamma_\parallel$ itself is a
generator of two-dimensional, Abelian rotations about ${\bf e}_{\bf k}$. It is not 
a normal 3-dimensional spin operator in momentum or real space.

Helicity states of good total angular momentum for a free photon can also be 
constructed 
\cite{Gottfried66}:
\begin{eqnarray} 
|kJM\lambda\rangle = \sqrt{\frac{2J+1}{4\pi}} 
\int d^2\hat{\bf k} D^{(J)}_{M\lambda}(\hat{\bf k})^*|{\bf k}\lambda\rangle,
\label{photonJMState}
\end{eqnarray}
where $\hat{\bf k} = {\bf e}_{\bf k}$, and $D^{(J)}_{M\lambda}$ is a 
representation matrix element of the finite rotation operator. Note that there 
are also four state labels ($k, J, M, \lambda$) here, and that a state of good 
$J$ does not have a preferred momentum direction ${\bf e}_{\bf k}$ or $\hat{\bf k}$.

All these well known expressions take such simple forms only
because the photon is free. If the $e\gamma$ interaction is also included, 
these expressions will become much more complicated.

\subsection{Photon multipole radiation}
\label{gamAMstates}

The helicity plane-wave state $|\alpha = {\bf k}\lambda\rangle$ of a free photon 
is also an eigenstate of 
$L^\gamma_\parallel = {\bf e}_{\bf k} \cdot {\bf L}^\gamma_{\rm phys} $ 
with eigenvalue $M_L = 0$, because the state is axially symmetric about {\bf k}. 
As such, it is one member of a set of states 
in circular coordinates with any integer eigenvalue $M_L$ called Laguerre-Gauss 
modes in optics \cite{Siegman86, Allen92}. The modes with $M_L \neq 0$ are 
physically observable helical waves \cite{Harris94} that do not concern us in 
this paper, but they do reinforce our understanding that the component 
$L^\gamma_\parallel$ of the orbital AM of a free photon along ${\bf e}_{\bf k}$ 
is measurable.

What about the remaining components of ${\bf L}^\gamma$? The issue can be addressed 
readily by using the elegant method of Berestetskii \cite{BLP71} for constructing 
coupled AM states $|(LS)JM_J\rangle$ or vector spherical harmonics that 
include the multipole radiation of massless free photons. The $m$-th component of 
any vector field, say in momentum space, can be written in the form
\begin{eqnarray} 
{\cal A}_m({\bf k}) = V_m f({\bf k}),
\label{VecField}
\end{eqnarray}
where a vector component $V_m$ can be separated from the remaining scalar field 
$f({\bf k})$ common to all components of $\bm{{\cal A}}$. Under rotation, the 
vector {\bf V} behaves just like the momentum {\bf k}. In fact, it can be
taken to be {\bf k} itself. The vector {\bf k} does not commute with the 
quantum orbital AM operator $\hat{\bf L} = - {\bf k} \times \hat{\bf x}$, but 
satisfies the commutation relations $[\hat{L}_j, k_m] = i\epsilon_{jmn}k_n$. 
Similar commutators therefore must hold for any vector {\bf V}:
\begin{eqnarray} 
[\hat{L}_j, V_m] = i\epsilon_{jmn}V_n.
\label{LVCommRels}
\end{eqnarray}
(The commutator is also valid in real space where the position vector plays the 
role of {\bf k} for the vector field. We use the same font for a quantity in 
both momentum and real spaces when its properties are very similar in both spaces.)

The photon orbital AM density appearing in the third equation of Eq. (\ref{Jgam}) 
then has the structure 
\begin{eqnarray} 
{\cal E}^*_m \hat{\bf L} (V_m f) 
&=& {\cal E}^*_m ( V_m\hat{\bf L} + {\bf e}_j i\epsilon_{jmn} V_n )f. 
\label{ELopA}
\end{eqnarray}
In matrix form, it reads 
\begin{eqnarray} 
\bm{{\cal E}}^\dag \hat{\bf L} ({\bf V} f) 
&=& \bm{{\cal E}}^\dag ({\bf V}\hat{\bf L} - \hat{\bf S} {\bf V} )f, 
\label{ELopAmatrix}
\end{eqnarray}
where {\bf V} is a column vector and $\bm{{\cal E}}^\dag$ is a row vector. 
One can then extract from it the Berestetskii identity as the operator equation 
in momentum or real space
\begin{eqnarray} 
\hat{\bf J} ({\bf V} f) = {\bf V}\hat{\bf L} f,
\label{ELopAeq}
\end{eqnarray}
where $\hat{\bf J} = \hat{\bf L} + \hat{\bf S}$ is the quantum total AM operator.
All these quantum AM operators satisfy the standard AM algebra:  
$\hat{\bf L} \times \hat{\bf L} = i\hat{\bf L},$ 
$\hat{\bf S} \times \hat{\bf S} = i\hat{\bf S},$ and 
$\hat{\bf J} \times \hat{\bf J} = i\hat{\bf J}.$

For the special case where the scalar field $f$ is the spherical harmonic $Y_{JM}$, 
one finds
\begin{eqnarray} 
\hat{\bf J}^2 ({\bf V} Y_{JM}) &=& {\bf V}\hat{\bf L}^2 Y_{JM} 
= J(J+1)({\bf V}Y_{JM}),
\nonumber \\
\hat{\bf J}_z ({\bf V} Y_{JM}) &=& {\bf V}\hat{\bf L}_z Y_{JM} 
= M({\bf V}Y_{JM}).
\label{JMeigenFn}
\end{eqnarray}
Hence the vector state $|{\bf V}Y_{JM}\rangle$ is a coupled $|(LS)JM\rangle$
state. The method actually works for any vector particle, massive as well as 
massless. For massive particles, one can show that the result is identical to 
that obtained from Clebsch-Gordan coefficients \cite{Blatt53}. 
 
For massless photons, however, the vector fields must reside on the physical 
plane. Hence the out-of-plane longitudinal vector fields ${\bf k}Y_{JM}$ 
are nonphysical and unrealizable. The permissible fields are two types of 
transverse fields often taken to be $\hat{\bf L}Y_{JM}$ and 
${\bf e}_{\bf x}Y_{JM}$. A transverse field of the first type is called a 
magnetic or MJ radiation, and may be denoted by the spectroscopic symbol 
$^3(L=J)_J$, with a single $L=J$ value for each $J$. A transverse field of the 
second type is an EJ radiation made up of a specific linear combination of the 
two states $^3(J-1)_J$ and $^3(J+1)_J$ of different $L$ values, a combination 
that is orthogonal to the linear combination found in the longitudinal field 
of the same $J$.

The unavailability of the longitudinal field for free photon states means that
it is not possible to construct the uncoupled $|LM_L\rangle |SM_S\rangle$
states from just the two transverse physical photon states. That is why one 
cannot measure simultaneously and exactly both $L$ and $M_L$ values of the 
photon orbital AM in free space. The photon $M_L$ is known to be zero along the 
momentum direction {\bf k}. The value $L=J$ for MJ radiations is known if $J$ is
known. On the other hand, each EJ radiation contains two possible $L$ values.
The average values of ${\bf L}^2$ and $M_L$ can be calculated if $J,M$ are
known, but not their exact values. So the orbital AM of a photon emitted by an
atom into free space is only partially observable.

More generally, there can be other complications to the quantum observability of 
the orbital AM even for massive matter particles such as electrons and nucleons 
when they are inside composite systems. For example, if $\hat{\bf L}$ does not 
commute with the Hamiltonian, as happens in the deuteron when a tensor 
nucleon-nucleon force is present, the relative orbital AM $\hat{\bf L}$ of an 
energy eigenstate such as the deuteron ground state is not a constant of motion 
(characterized by simultaneous quantum numbers) of the quantum energy eigenstate. 
For this reason, the deuteron D-state probability 
$p_D = \langle \hat{\bf L}^2 \rangle/6\hbar^2$ is not the eigenvalue of a physical 
observable that commutes with the Hamiltonian. However, it can be indirectly deduced 
if somehow we can measure $\langle \hat{\bf L}^2 \rangle$. 

A final remark:  For quantization in a covariant gauge, the longitudinal and 
time-like photons are both unphysical and not directly observable as on-shell 
photons. However, they do give rise to observable effects such as the Coulomb 
interaction among static charges \cite{Wong10} and Coulomb 
excitations/deexcitations of atomic and nuclear states. However, these effects are 
also correctly described for quantization in the Coulomb gauge using only 
transverse photons.

\section{Gauge invariant operators in QCD}

In QED, the gauge dependent and nondynamical part ${\cal A}_{\rm nd}$ of the 
vector potential ${\cal A}^\mu$ resides in the 2-dimensional energy-momentum (03) 
subspace orthogonal to the physical (12)-plane that contains the gauge-independent 
part $\bm{{\cal A}}_\perp$. In QCD, one intuitively expects the same separation 
because color is an internal attribute independent of spacetime itself. We 
obtain in Sect. A an explicit perturbative solution for the nondynamical 
${\cal A}_{\rm nd}$ and find that it is indeed confined to the 
(03)-subspace. In Sect. B, both nondynamical and dynamical momentum
operators are found to have appearances similar to the QED operators. However, 
the color electric field in the gluon operators differs from the QED field in 
an extra term quadratic in $A$ and proportional to the QCD coupling constant $g$. 
This term describes an interaction that changes the value of the gluon momentum
itself. The same nonlinear electric field term is found in Sect. C to add 
nonlinear interaction structures to our orbital AM and total AM operators 
in QCD that are otherwise nondynamical.

\subsection{Nondynamical vector potentials in classical CD}

The color Maxwell field is $F^{\mu\nu} = F^{\mu\nu}_a T_a$, where
\begin{equation}
F^{\mu\nu}_a = \partial^{\mu}A^{\nu}_a - \partial^{\nu}A^{\mu}_a
+ g C_{abc} A^{\mu}_bA^{\nu}_c.
\label{QCDfield}
\end{equation}
The color operators $T_a$ are $3 \times 3$ matrices in color space that satisfy 
the $SU_3$ commutation relations
\begin{equation}
\left[T_a, T_b\right] = iC_{abc} T_c,
\label{SU3CommRel}
\end{equation}
where the structure constant $C_{abc}$ is totally antisymmetric in its indices.
The nondynamical (nd) vector potential 
$A^{\mu}_{\rm nd} =A^\mu_{{\rm nd}\, a}T_a$ of the gluon field is defined by 
the condition that it generates no color field:
\begin{equation}
F^{\mu\nu}_{{\rm nd}\, a} = \partial^{\mu}A^{\nu}_a-\partial^{\nu}
+ g C_{abc} A^{\mu}_bA^{\nu}_c = 0,
\label{QCDpureCondition}
\end{equation}
where the subscript ``nd'' has been dropped from $A^{\mu}_a$ for notational 
simplicity. We do not know any analytic solution of this nonlinear differential 
equation for $A^{\mu}_a$, but look instead for a perturbative solution in powers 
of the QCD coupling constant $g$:
\begin{equation}
A^{\mu}_a = \sum_{n=0}^\infty g^n A^{(n)\mu}_a.
\label{Apert}
\end{equation}
The first term $A^{(0)\mu}_a$ is just the linear QED solution, because it 
satisfies the QED zero-field condition 
$\partial^{\mu}A^{\nu}_a-\partial^{\nu}A^{\mu}_a = 0$. The remaining terms are
solutions of the differential equations
\begin{equation}
\partial^\mu A^{(n)\nu}_a - \partial^\nu A^{(n)\mu}_a
= C_{abc} \sum_{j=0}^{n-1} A^{(j)\mu}_b A^{(n-1-j)\nu}_c.
\label{Apertn}
\end{equation}

In the Fourier space $k^\mu =(\omega, 0, 0, |{\bf k}|)$, Eq.(\ref{Apertn}) 
reduces to the matrix equation 
\begin{eqnarray}
{\cal L}^{(n)\mu\nu}_a &\equiv& 
k^\mu{\cal A}^{(n)\nu}_a - k^\nu {\cal A}^{(n)\mu}_a \nonumber \\
&=& {\cal R}_a^{(n-1)\mu\nu},
\label{matrixEqPure}
\end{eqnarray}
for $n > 0$. Here
\begin{eqnarray}
{\cal R}_a^{(n-1)\mu\nu} 
&=& - iC_{abc} \sum_{j=0}^{n-1} 
\bm{(} {\cal A}^{(j)\mu}_b \ast {\cal A}^{(n-1-j)\nu}_c \bm{)},
\nonumber \\
\bm{(} {\cal A}^{(j)\mu}_b \ast {\cal A}^{(n-1-j)\nu}_c \bm{)}
&=& \frac{1}{(2\pi)^4} 
\int {\cal A}^{(j)\mu}_b(k-k') \nonumber \\
&& \qquad \times \; {\cal A}^{(n-1-j)\nu}_c(k') d^4k'.
\label{rhsRn-1}
\end{eqnarray}
have been obtained from the convolution theorem. Note that 
${\cal L}^{(0)\mu\nu}_a = {\cal F}^{\mu\nu}_{{\rm nd}\, a} = 0$.

Equation (\ref{matrixEqPure}) equates two $4 \times 4$ antisymmetric matrices. 
On the right, the matrix ${\cal R}_a^{(n-1)\mu\nu}$ will be shown inductively to 
have only two nonzero matrix elements, at positions $\mu\nu = 30, 03$ where 
${\bf e}_3 = {\bf e}_{\bf k}$. On the left, the matrix ${\cal L}^{(n)}$ has the 
structure of a general antisymmetric field tensor with 6 zero matrix elements, 
4 on the diagonal and 2 at positions 12 and 21 (because 
$k^1 = k^2 = 0$). The remaining 10 nontrivial matrix 
elements are in 5 antisymmetric pairs, 3 electric field components and 2 
magnetic induction components. The component $B_\parallel = B_3$ vanishes
in this QED-like field tensor because magnetic monopoles are absent in QED. 

With the matrices written in the Cartesian order $(3, 0; 1, 2)$, the 4 matrix 
elements in one of the two $2 \times 2$ off-diagonal block submatrices is
assumed temporarily to satisfy the algebraic equations
\begin{eqnarray}
{\cal L}^{(n)\beta\sigma}_a = - k^\sigma {\cal A}^{(n)\beta}_a 
= {\cal R}^{(n-1)\beta\sigma}_a = 0,
\label{EqOffDiag}
\end{eqnarray}
where $\beta = 1,2, \;\sigma = 3,0$. This assumption, that 
${\cal R}_a^{(n-1)\beta\sigma} = 0$, will be justified by induction later. 
The unique solutions of Eq. (\ref{EqOffDiag}) are then
${\cal A}^{(n)\beta}_a = 0$, for $\beta = 1,2$. In other words, the temporary
assumption is that the driving term ${\cal R}^{(n-1)\beta\sigma}_a$ does not 
have any block off-diagonal matrix element to drive ${\cal A}^{(n)\beta}_a$ 
onto the physical (12)-plane. In this way, the nondynamical ${\cal A}_a$
stays in the (03)-subspace orthogonal to the physical plane. The same solution 
can be obtained from the other off-diagonal block submatrix.

The 5th and last equation is in the (30) subspace alone and reads
\begin{eqnarray}
{\cal L}^{(n)30}_a = |{\bf k}| {\cal A}^{(n)0}_a - \omega {\cal A}^{(n)3}_a
 = {\cal R}^{(n-1)30}_a. 
\label{Eq03Diag}
\end{eqnarray}
A second equation is needed to solve uniquely for the two unknowns in the
$n$-th order. This is provided by choosing a gauge, but so far we have considered
only the velocity gauge condition (\ref{gaugeCond}) for QED that is good in 
QCD only for order $n=0$. For $n>0$, one can in principle choose any velocity
gauge in every order $n$ in a totally random manner, but such a choice gives 
infinitely many possibilities that cannot even be written down explicitly. The 
simplest choice is the opposite extreme, namely, to use the same velocity gauge 
in every order of $g$:
\begin{eqnarray}
{\cal A}^{(n)3}_a = s\frac{\alpha\omega}{|{\bf k}|}{\cal A}^{(n)0}_a,
\label{VelGaugeAnyOrder}
\end{eqnarray}
where $s = \pm 1$ and the gauge parameter $\alpha = 1/v_g^2$ is defined by the 
gauge velocity $v_g$.

The complete zeroth-order nondynamical vector potential [in the component order 
$(3, 0; 1, 2)$] is known from QED to be:
\begin{eqnarray}
\bm{{\cal A}}^{(0)}_a = ({\cal A}^{(0)3}_a,{\cal A}^{(0)0}_a; 0, 0)
\label{Aorder0}
\end{eqnarray}
for every color $a$. The antisymmetric matrix ${\cal R}^{(0)}_a$ thus have two 
rows and two columns of zeros where the matrix index is on the physical plane, 
i.e., $\mu = 1,$ and 2. Hence it has only two nonzero matrix elements, at 
positions 30 and 03. This verifies the assumed property of ${\cal R}^{(n)}_a$ 
for $n=0$.

The first-order equation (\ref{Eq03Diag}) (with $n=1$) and the order-independent
gauge condition (\ref{VelGaugeAnyOrder}) can now be solved for the two unknowns. 
The result,
\begin{eqnarray}
{\cal A}^{(1)0}_a &=& \frac{|{\bf k}|}{|{\bf k}|^2 - s\alpha\omega^2} {\cal R}^{(0)30}_a, 
\nonumber \\  
{\cal A}^{(1)3}_a &=& s\frac{\alpha\omega}{|{\bf k}|}{\cal A}^{(1)0}_a,
\label{SolnFirstOrder}
\end{eqnarray}
is a vector confined to the (30)-plane, just like $\bm{{\cal A}}^{(0)}_a$.
The antisymmetric matrix ${\cal R}^{(1)}_a$ constructed from it also vanishes 
everywhere except at positions 30 and 03. The second-order solution can then be 
obtained in exactly the same way. We can thus bootstrap our way up the 
perturbative expansion to give the complete solution
\begin{eqnarray}
{\cal A}^0_{{\rm nd}, a} &=& \frac{|{\bf k}|}{|{\bf k}|^2 - s\alpha\omega^2} 
{\cal R}^{30}_a, 
\nonumber \\
{\cal A}^3_{{\rm nd}, a} &=& s\frac{\alpha\omega}{|{\bf k}|}
{\cal A}^0_{{\rm nd}, a}, 
\nonumber \\
{\cal R}^{30}_a &=& \sum_{n=0}^\infty g^n {\cal R}^{(n)30}_a,
\label{SolnAllOrders}
\end{eqnarray}
confined to the (30)-plane. The nonlinear dynamics changes the actual
functional form on this plane, but preserves unchanged its avoidance of the 
physical (12)-plane.

Any color vector potential in a chosen Lorentz frame can thus be written as
$A^\mu = A^\mu_{\rm quan} + A^\mu_{\rm nd}$, where $A^\mu_{\rm quan}$ is
its dynamical part in the gauge where second quantization is realized. 
Here the color label is unimportant and is omitted. For canonical quantization 
in the Coulomb or physical gauge, we have 
$A^\mu_{\rm quan} = A^\mu_{\rm phys} = A^\mu_C$. With the same velocity gauge 
condition as QED, it follows that ${\bf A}_{\rm phys} = {\bf A}_C = {\bf A}_\perp$ 
holds in any color.  

The use of the Coulomb gauge also simplifies the nonlinear (NL) part of the 
field tensor. With $A_\parallel = A^3 = 0$, $F^{\mu\nu}_{\rm phys\,NL}$ is a 
$4 \times 4$ matrix that is not only 
antisymmetric, but is also zero in the row and column involving a $j=3$ 
component. Hence it contains only 6 nonzero matrix elements or 3 nonzero field 
components: $B^3, E^1$ and $E^2$. The presence of $B^3 \neq 0$ means that the 
nonlinear color dynamics generates color magnetic monopoles. As a result, the 
CCD momentum density contains additional terms including one 
$ {\bf E}_{\rm NL} \times {\bf B}_{\rm NL} 
= {\bf E}_{\perp{\rm NL}} \times {\bf B}_{\parallel{\rm NL}}$ that gives
a contribution in a direction perpendicular to {\bf k}. Similar complications 
arise in the AM operators.

A final note: Since the differential equation (\ref{QCDpureCondition}) has a 
driving term quadratic in $A^\mu$, it seems likely that the equation has another 
solution.  However, this cannot be a perturbative solution in powers of $g$ 
because the perturbative solution obtained here is the unique perturbative 
solution for the chosen gauge. If the second solution exists, it 
will be a nonperturbative solution. 

While it is interesting that QED-like gauge conditions such as 
Eq. (\ref{VelGaugeAnyOrder}) also works in QCD, it is worth noting that the 
formalism developed here works for more complicated gauge relations as well.
In the rest of the paper, we shall consider only the QED-like perturbative 
solutions obtained in this subsection.

\subsection{Momentum operators in QCD}

The color Maxwell field (\ref{QCDfield}) is conveniently written as two 3-vector 
fields
\begin{eqnarray}
{\bf E}_a &=& - \partial_t{\bf A}_a - \bm{\nabla}A^0_a 
    - gC_{abc} A^0_b{\bf A}_c, 
\label{QCDE} \\
{\bf B}_a &=&  \bm{\nabla} \times {\bf A}_a 
    - \frac{1}{2} gC_{abc} \left( {\bf A}_b \times {\bf A}_c \right).
\label{QCDB}
\end{eqnarray}
Of the color Maxwell equations 
$\partial_{\mu}F^{\mu\nu}_a + g C_{abc} A_{\mu,b}F^{\mu\nu}_c = J^\nu_a$,
we shall only need the color Gauss law:
\begin{eqnarray}
\bm{\nabla} \cdot {\bf E}_a 
&=& \rho_a + gC_{abc} {\bf A}_b \cdot {\bf E}_c,
\label{colorGauss}
\end{eqnarray}
where in QCD, $\rho_a = g \psi^{\dag}T_a\psi$.

The derivation of momentum and AM operators in QCD differs from that already 
presented for QED only by the addition of nonlinear terms proportional to $g$. 
The similarity allows the separation 
${\bf P}^g = {\bf P}^g_{\rm phys} + {\bf P}^g_{\rm pure}$ to be made in any 
velocity gauge when quantization is realized on the physical (12)-plane 
perpendicular to ${\bf e}_{\bf k}$. Again ${\bf P}^g_{\rm pure}$ can be 
simplified by using the QCD version of the vector identity (\ref{vectorID}) by 
considering the special case ${\bf B}_{{\rm pure}\, a} = 0$, for which 
${\bf A}_{{\rm pure}\, a} = {\bf A}_{\parallel a}$ and therefore both terms
on the right of Eq. (\ref{QCDB}) vanish separately. Note that the nonlinear 
term on the right of Eq. (\ref{QCDB}) is nonzero only when the spatial 
direction of the vector potential ${\bf A}_b$ depends on the color label
$b$. We do not know if such solutions with space-color correlations exist. 

Using ${\bf E} = {\bf E}_{\parallel a}$, we find with the implied sum over the 
color label $a$
\begin{eqnarray}
0 = E_{\parallel a} ( \bm{\nabla}A_{\parallel a} ) 
+ ( \bm{\nabla} \cdot {\bf E}_{\parallel a} ) {\bf A}_{\parallel a} 
- \nabla^j ( E^j_{\parallel a}{\bf A}_{\parallel a} ).
\label{QCDvectorID}
\end{eqnarray}
The color Gauss law (\ref{colorGauss}) has an additional nonlinear term on the 
right, but its contribution to Eq. (\ref{QCDvectorID}) on the right is 
$gC_{abc} {\bf A}_{\parallel a}{\bf A}_{\parallel b}{\bf E}_{\parallel c} = 0$. 
Hence all the QED decompositions and rearrangements can be taken over 
wholesale by substitutions of the type 
$e{\bf A}^j \rightarrow g{\bf A}^j = g{\bf A}^j_aT_a$
and by summing over an additional color label where appropriate.

For example, the manifestly gauge-invariant, nondynamical quark momentum
operator takes the same form as Eq. (\ref{PeGauInv})
\begin{eqnarray}
{\bf P}^q_{\rm nd} &=& {\bf P}^q + {\bf P}^g_{\rm pure}  \nonumber \\
&=& \int \psi^{\dag} (\hat{\bf p} - g{\bf A}_\parallel)\psi \,d^3x
\label{PqGauInv}
\end{eqnarray}
when the color content is not made explicit. 
However, the color structure has to be made explicit to display explicit gauge 
invariance: The quark field $\psi$ is a column vector in color space. Under a 
nondynamical gauge transformation, it is changed to $\psi' = U\psi$, where 
$U = \exp{(i\lambda)}$ is a $3 \times 3$ matrix in color space because 
$\lambda = \lambda_aT_a$. 
The quark momentum density $\psi^\dag (\hat{p}^\mu - eA^\mu_{\rm pure}) \psi$ 
then remains unchanged if $(\hat{p}^\mu - eA^\mu_{\rm pure})' = U(\hat{p}^\mu - eA^\mu_{\rm pure})U^\dag$. Just as in QED, both $k^\mu$ and 
${\cal A}^\mu_{\rm pure}$ have only nonzero $\mu = 0, 3$ components in momentum 
space. Their identical mathematical structures explain why they appear naturally 
together. The density $\psi^\dag {\bf A}_\perp \psi$ in the dynamical operators 
of Ji \cite{Ji97} is gauge invariant if ${\bf A}_\perp' = U {\bf A}_\perp U^\dag$.

With ${\bf P}^g_{\rm pure}$ removed, the remaining gluon momentum density 
$E^j_a( \bm{\nabla} A^j_a ) - E^j_{\parallel a}
( \bm{\nabla} A^j_{\parallel a} ) $ can be written as three terms of spatial 
structures $\perp\parallel, \parallel\perp,$ and $\perp\perp$. The two cross 
terms vanish separately, leaving
\begin{eqnarray}
{\bf P}^g_{\rm phys} &=&  {\bf P}^g_{\rm phys\, L} + {\bf P}^g_{\rm phys\, NL},
\nonumber \\
{\bf P}^g_{\rm phys\, C} 
&=& \int E^j_{\perp a{\rm C}}( \bm{\nabla} A^j_{\perp a} ) \,d^3x, 
\quad _{\rm C = L, NL}.
\label{PgPhys}
\end{eqnarray}
The linear (L) and nonlinear (NL) terms come from the first and second terms,
respectively, of the color electric field on the physical plane
\begin{eqnarray}
{\bf E}_{\perp a} &=& - \partial_t{\bf A}_{\perp a}  
    - gC_{abc} A^0_b{\bf A}_{\perp c}, \nonumber \\
&=&   {\bf E}_{\perp a{\rm L}} + {\bf E}_{\perp a{\rm NL}}.   
\label{QCDEperp} 
\end{eqnarray}
The NL term is an interaction term proportional to $g$ that is not present in QED.
It contains the nonlinear momentum density in momentum space of
\begin{eqnarray}
- gC_{abc} \bm{(} {\cal A}^0_b \ast {\cal A}^j_{\perp c} \bm{)}({\bf k})
[i{\bf k} {\cal A}^j_{\perp a}({\bf k})]  \neq 0.
\label{PgPhysNLden}
\end{eqnarray}
If second quantization is still defined by {\bf A} of Eq. (\ref{quantAperp}) for 
each color, ${\bf E}_{\perp a}$ remains linear in ${\bf a}$ and ${\bf a}^\dag$,
but the interaction contained in the nonlinear term gives rise to gluons of 
arbitrary momentum ${\bf k}'$ inside the convolution. [The convolution contributes 
to Eq. (\ref{PgPhysNLden}) except when ${\bf k}' ={\bf k}$.] 
Hence gluon momenta are more complicated than photon momenta.

\subsection{Angular momentum operators in QCD}

The construction of AM operators proceeds similarly to QED. 
The details will also be given in the Appendix. The results are summarized here.

The expression obtained directly from the QCD Lagrangian is
\begin{eqnarray}
{\bf J} &=& {\bf S}^q + {\bf L}^q + {\bf S}^g + {\bf L}^g \nonumber \\
&=& \int \psi^{\dag} (\bm{\Sigma}/2) \psi \,d^3x
+ \int \psi^{\dag} \hat{\bf L} \psi \,d^3x   \nonumber \\
&&+\; \int {\bf E}_a \times {\bf A}_a \,d^3x 
+ \int  E^j_a (i\hat{\bf L} A^j_a) \,d^3x, 
\label{QCDJtot}
\end{eqnarray}
The gluon operators can be re-written as  
\begin{eqnarray}
{\bf S}^g + {\bf L}^g 
&=& {\bf S}^g_\parallel + {\bf L}^g_{\rm phys} 
+ {\bf L}^g_{\rm pure}, \nonumber \\
{\bf S}^g_\parallel &=&  \int {\bf E}_{\perp a} \times {\bf A}_{\perp a} \,d^3x, 
\nonumber \\
{\bf L}^g_{\rm phys} &=& \int  E^j_{\perp a} (i\hat{\bf L} A^j_{\perp a}) \,d^3x,
\nonumber \\
{\bf L}^g_{\rm pure} &=& 
-\int \psi^{\dag} ({\bf x} \times g{\bf A}_\parallel) \psi \,d^3x.
\label{QCDJg}
\end{eqnarray}
We get two physical, gauge-invariant terms ${\bf S}^g_\parallel$
and ${\bf L}^g_{\rm phys}$, and a nondynamical gauge-dependent term to be
added to the quark orbital AM to make it manifestly gauge invariant: 
\begin{eqnarray}
{\bf L}^q_{\rm nd} &=& {\bf L}^q + {\bf L}^g_{\rm pure}  \nonumber \\
&=& \int \psi^{\dag} {\bf x} \times (\hat{\bf p} - g{\bf A}_\parallel) \psi \,d^3x. 
\label{qLnd}
\end{eqnarray}

Ji's dynamical quark and gluon AM operators can be obtained from the electron 
and photon operators by a simple change of names when the color index is not 
shown explicitly.

Returning to our nondynamical gluon operators (leaving out the pure gauge term
${\bf L}^g_{\rm pure}$ for simplicity because it is not involved in the 
following discussion), we note that the following results for their nonlinear terms:
\begin{eqnarray}
{\bf S}^g_{\parallel {\rm NL}} &=&  
- gC_{abc} \int (A^0_b {\bf A}_{\perp c}) \times {\bf A}_{\perp a} \,d^3x = 0, 
\label{SgNL}  \\
{\bf L}^g_{\rm phys\,NL} &=& 
- gC_{abc} \int  (A^0_b A^j_{\perp c}) i\hat{\bf L} A^j_{\perp a} \,d^3x \neq 0.
\label{LgNL}
\end{eqnarray}
That is, ${\bf S}^g_{\parallel {\rm NL}}$ in Eq. (\ref{SgNL}) vanishes for all 
solutions ${\bf A}_{\perp a}$ whose direction in real space {\bf x} is the same 
for all colors. In Eq. (\ref{LgNL}), the color sum would have given nothing if 
it were of the form $C_{abc}X_bX_c = 0$ with two identical and equivalent vectors 
$X_a$ and $X_b$ that can be interchanged freely by re-labeling the color indices. 
However, the actual expression involves $A^j_{\perp c}$ and 
$\hat{\bf L} A^j_{\perp a}$ that are not identical vectors. 
It is therefore nonzero.

A more general result can be derived for the sum ${\bf J}^g_{\rm phys\,NL} 
= {\bf L}^g_{\rm phys\,NL} + {\bf S}^g_{\parallel {\rm NL}}$ of 
Eqs. (\ref{SgNL}, \ref{LgNL}). The vector form of the Berestetskii identity 
Eq. (\ref{ELopAvector}) can be written in real space for 
${\bf A}_{\perp a}({\bf x}) = {\bf V}_\perp f_a({\bf x})$, where $f_a({\bf x})$ is
a scalar function:
\begin{eqnarray} 
E^j_{\perp a} i\hat{\bf L} (V^j_\perp\ f_a) 
+ {\bf E}_{\perp a} \times {\bf A}_{\perp a}
= ({\bf E}_{\perp a} \cdot {\bf V}_\perp) i\hat{\bf L}f_a. 
\label{BerestetskiiVectorID}
\end{eqnarray}
Hence
\begin{eqnarray}
{\bf J}^g_{\rm phys\,NL} &=& 
- gC_{abc} \int  (A^0_b f_c i\hat{\bf L} f_a ) {\bf V}^2_\perp \,d^3x 
\neq 0,
\label{JgNL}
\end{eqnarray}
because $f_c$ and $\hat{\bf L} f_a$ are different functions in general. 
For example, if $f = Y_{LM}$, then $\hat{\bf L} f$ can be written as a sum of
three terms, each proportional to  ${\bf e}_z Y_{LM}$, ${\bf e}_+ Y_{LM-1}$ 
and ${\bf e}_- Y_{LM+1}$, respectively. The first term contributes nothing to 
Eq. (\ref{JgNL}), but the remaining two terms do contribute.
Thus ${\bf J}^g_{\rm phys}$ also has a more complicated structure than 
${\bf J}^\gamma_{\rm phys}$ in QED.

\section{Concluding remarks}

Ji in his published Comment \cite{Ji10} on our recent paper \cite{Chen08} on 
angular momentum (AM) operators in gauge theories has expressed a number of 
concerns. First, he claims that our ``new notion of gauge invariance clashes 
with locality and Lorentz symmetry.'' Let us first respond to his concern about 
Lorentz symmetry. As we have explained in the Introduction, manifest Lorentz 
covariance and gauge invariance can coexist explicitly only in a covariant gauge 
such as the Lorenz gauge. For quantization in a non-covariant gauge, we need to 
retrace our steps by transforming to a covariant gauge in the chosen Lorentz frame 
and use the latter's Lorentz covariance to restore explicitly Lorentz symmetry. 
The point is that once Lorentz covariance is established initially, it does not 
have to be maintained in subsequent steps. So there is no need for us to look at 
the vector potential in another Lorentz frame when we quantize a gauge-invariant
theory in a non-covariant gauge.

However, we find hidden Lorentz symmetries in our formalism in Sect. 
\ref{AinCED}. First, $\bm{{\cal A}}_\perp^2$ is a Lorentz scalar, where
$\bm{{\cal A}}_\perp$ is the gauge-independent part of {\it any} vector potential 
$\bm{{\cal A}}$ in our chosen family of gauges in 4-momentum space, the part 
that lies on the 2-dimensional physical plane perpendicular to the photon/gluon 
momentum {\bf k}. This is an interesting and desirable feature, even though it 
does not involve the whole vector potential. 

Ji's concerns about both Lorentz symmetry and locality are also resolved
by our choice to second quantize only on the physical plane. We have shown 
explicitly in Sect. \ref{CanonQuantization} that for any choice of gauge among
our chosen family, second quantization nevertheless proceeds like a standard 
canonical quantization in the Coulomb gauge for the gauge-independent part 
$\bm{{\cal A}}_{\rm phys} \equiv \bm{{\cal A}}_\perp$. No unusual or unacceptable 
nonlocal operators appear. The simple Coulomb-gauge structure is always obtained 
because we do not use the complete vector potential ${\cal A}^\mu$ directly. 
Ji's concern may apply to quantization for the complete ${\cal A}^\mu$, 
a procedure we do not recommend. Moreover, our quantization procedure is 
mathematically identical to Coulomb-gauge quantization and shares the latter's 
implicit Lorentz-invariant character in the resulting quantum field theories 
\cite{Zumino60, Lee81}. 

Thirdly, we have taken considerable space in the Introduction to explain why
the nondynamical gauge transformation in the vector potential and the associated
path-independent phase factor in the fermion field contain the essence of the
concept of a gauge invariance that does not change the assumed dynamics of the
problem. After this demonstration of gauge invariance, calculations can then be 
made in any convenient gauge. In Sect. \ref{QEDmom}, we have pointed out some 
theoretical and practical advantages of not including dynamics in the definition 
of operators of certain wholly or partially observable physical attributes, 
at least in QED. 

However, there is a different gauge principle, one that gives rise to the local 
gauge theory of interactions, where dynamics can be generated by the use of 
dynamical, path-dependent or nonintegrable phase factors that multiply into a 
dynamics-free fermion expression. From our perspective, nothing forbids the use 
of the dynamical operators advocated by Ji and others. We have described in this 
paper, especially Sect. IIB, the advantages of our nondynamical operators based 
on $\hat{p}^\mu - gA^\mu_{\rm pure}$ over Ji's dynamical operators based on
$\hat{p}^\mu - gA^\mu$. Even the mathematical structure of our operators is 
simpler because in 4-momentum space, both $k^\mu$ and ${\cal A}^\mu_{\rm pure}$ 
have nonzero components only for $\mu = 0, 3 \,({\rm or} \parallel)$, i.e.,
in the subspace orthogonal to the physical plane. Which type of 
operators will be more convenient to use will be settled by future practices. 

We have also pointed out in Sect. \ref{QEDmom} an essential difference between
classical and quantum mechanics in their dynamical descriptions. The dynamical 
momentum appears naturally in classical particle dynamics where a point particle 
carries around its full dynamics as it moves in space. In contrast, an observed 
quantum particle is nondynamical in the sense that it is on the energy shell and 
does not contain any information on its past dynamical history in the quantum 
source system where it existed before the detection. One has to measure many such 
particles from the same quantum source in order to deduce their common dynamics 
from their measured distribution functions. 

Among the issues under popular discussion is the question of whether the orbital
AM and spin of photons and gluons are separately observable. For direct 
measurements on free photons, the answer is unambiguous, as we have described in 
Sect. \ref{gamAMstates}. The transverse photon states can be expressed in terms of 
one or two states of the type $|(LS)JM\rangle$, where $S, J, M$ are good quantum 
numbers. In MJ photons, only one state $L=J$ appears, whereas in EJ photons, two 
states with $L = J\pm 1$ appear in a known linear combination. Hence $S, J, M$ are
completely observable, and $L$ is observable for MJ photons and only partially
observable for EJ photons. In these $|(LS)JM\rangle$ states,  $M_L$ or $M_S$ is 
not separately observable, but the combination $M = M_L + M_S$ is observable.
However, in states of good momentum {\bf k}, $M_L = 0$ and $M_S = \lambda = \pm 1$ 
along ${\bf e}_{\bf k}$ are separately observable.

For gluons, however, our analysis shows that even our nominally nondynamical {\bf P} 
and {\bf J} operators are not free of interaction contributions. This is because 
the color electric field ${\bf E}_\perp$ appearing in them contains a term 
quadratic in the vector potential $A^\mu$ and proportional to the color coupling 
constant $g$, as shown explicitly in Eq. (\ref{QCDE}). The fact that gluons are 
always confined and never free can cause additional complications. The present 
status of, and open problems in, the experimental measurement of quark and gluon contributions to the proton spin have been reviewed recently by Bass \cite{Bass05}.

Finally, Ji \cite{Ji10} has pointed out the energy scale dependence of various
quantities involved in the nucleon spin problem. This dependence comes from the 
fact that all physics, including the strong coupling constant 
$\alpha_s(\mu) = g^2(\mu)/4\pi$, is a function of the energy scale $\mu$ at which
the measurement is made. He has emphasized that even the quark spin, which has 
gluon contributions from the Adler-Bell-Jackiw triangle anomaly, 
is not free of this complication. Indeed, as discussed in the last paragraph,
all QCD momentum and angular momentum operators are $g$-dependent.
The lesson here seems to be that theoretical analyses should match the energy 
scale of the experimental data. However, various commutation relations that 
simplify calculations at one energy scale can still be used in spite of their 
intrinsic scale dependence. Again, the question will be settled by future practices.

In contrast, the gauge invariance addressed in this paper is a relatively simple 
issue. The simple answer given here is that this gauge invariance is actually a 
nondynamical concept in both QED and QCD.

\appendix

\section{Angular momentum operators in QED and QCD}

In this Appendix, we show how the AM operators calculated from the Lagrangian in 
QED and QCD are reduced to a form suitable for canonical quantization in the 
Coulomb gauge.

For QED, Eq. (\ref{Jgam}) is obtained from its parent expression (\ref{Jtot}) if 
their difference (expressible as two groups of terms) vanishes. The first group of 
terms is
\begin{eqnarray}
\int  E^j ({\bf x} \times \bm{\nabla} A^j_\perp) \,d^3x 
&+& \int \psi^{\dag} ({\bf x} \times e{\bf A}_\parallel) \psi \,d^3x  \nonumber \\
&+& \int {\bf E} \times {\bf A}_\perp \,d^3x = 0.
\label{JgamApara}
\end{eqnarray}
This equation can be derived by using the differential vector identity 
(\ref{angMomDenID}) for the special case ${\bf B} = {\bf B}_{\rm pure} 
= {\bf B}_\parallel = 0, \;{\bf A} = {\bf A}_\parallel$:
\begin{eqnarray}
0 &=& E^j ({\bf x}\times\bm{\nabla}) A^j_\parallel 
+ (\bm{\nabla}\cdot{\bf E})({\bf x}\times{\bf A}_\parallel) \nonumber \\
&& +\; {\bf E} \times {\bf A}_\parallel
 - \nabla^j[E^j({\bf x}\times{\bf A}_\parallel)].
\label{angMomDenIDpara}
\end{eqnarray}
The spatial integral of the last term on the right vanishes when it is changed 
into a surface integral at $\infty$ where the integrand vanishes. 
Equation (\ref{JgamApara}) then follows with the help of the Gauss law
$\bm{\nabla} \cdot {\bf E} = \rho = e\psi^{\dag}\psi$.

The second group of terms is more transparent when written as a momentum 
integral 
\begin{eqnarray}
\int  [ {\cal E}^{j*}_\parallel (i\hat{\bf L} {\cal A}^j_\perp)
+ \bm{{\cal E}}^*_\parallel \times \bm{{\cal A}}_\perp ] \,d^3k = 0. 
\label{JgamEperp}
\end{eqnarray}
To verify its vanishing value, we write the Berestetskii identity (\ref{ELopAmatrix}) 
for two arbitrary vectors $\bm{{\cal E}}$ and $\bm{{\cal A}} = {\bf V}f$, 
where $f = f({\bf k})$ is a scalar field, in the alternative vector form
\begin{eqnarray} 
{\cal E}^{j*} \hat{\bf L} (V^j f) 
= (\bm{{\cal E}}^* \cdot {\bf V}) \hat{\bf L}f 
+ i\bm{{\cal E}}^* \times \bm{{\cal A}}. 
\label{ELopAvector}
\end{eqnarray}
For the special case $\bm{{\cal E}} = \bm{{\cal E}}_\parallel, \;
\bm{{\cal A}} = \bm{{\cal A}}_\perp$, the identity shows that the integrand in 
Eq. (\ref{JgamEperp}) is equal to the 
$\bm{{\cal E}}_\parallel^* \cdot {\bf V}_\perp$ term which vanishes at every value 
of {\bf k}. This completes the derivation of Eq. (\ref{Jgam}) from 
Eq. (\ref{Jtot}).

For QCD, the second group of terms [the left side of Eq. (\ref{JgamEperp})] also 
contribute nothing because Eq. (\ref{ELopAvector}) holds for each color. 
For the first group of terms [the left side of Eq. (\ref{JgamApara})], we
need to use the differential vector identity (\ref{angMomDenIDpara}) with an
implied sum over the color index $a$:
\begin{eqnarray}
0 &=& E^j_a ({\bf x}\times\bm{\nabla}) A^j_{\parallel a}
+ (\bm{\nabla}\cdot{\bf E}_a)({\bf x}\times{\bf A}_{\parallel a}) \nonumber \\
&& + \; {\bf E}_a \times {\bf A}_{\parallel a}
 - \nabla^j[E^j_a( {\bf x}\times{\bf A}_{\parallel a} )].
\label{QCDangMomDenIDpara}
\end{eqnarray}
The only real difference from the linear QED structure is the additional 
contribution coming from the nonlinear term in the color Gauss law 
(\ref{colorGauss}). With the implicit sum over the color index $a$, this 
nonlinear contribution vanishes because
\begin{eqnarray}
&& gC_{abc} ({\bf A}_{\parallel b} \cdot {\bf E}_c )
({\bf x} \times {\bf A}_{\parallel a})  \nonumber \\
&=& g{\bf e}^j \epsilon_{jm3} x^m 
(C_{abc} A^3_a A^3_b E^3_c ) = 0,
\label{NLGaussTerm}
\end{eqnarray}
where ${\bf e}_\parallel = {\bf e}_3$. Hence the first group of terms 
vanishes also for QCD. Thus Eq. (\ref{QCDJg}) follows from Eq. (\ref{QCDJtot}).

\end{document}